\documentclass[12pt,tightenlines,nofootinbib,amsmath,amssymb,aps,prd,superscriptaddress,notitlepage]{revtex4-1}

\usepackage{graphicx}
\usepackage[caption=false]{subfig}

\usepackage{amsmath,amssymb}
\usepackage{amsfonts,amssymb,mathrsfs}

\usepackage[bookmarks=false,pdfstartview=FitH]{hyperref}
\usepackage[all]{hypcap}

\usepackage[left=1.1in,right=1.1in]{geometry}

\def\be{\begin{equation}}
\def\ee{\end{equation}}
\def\nn{\nonumber}
\def\f{\frac}

\def\pl{{\rm Pl}}
\def\lp{\ell_\pl}
\def\b{\bar}

\def\h{\hat}
\def\t{\tilde}
\def\wh{\widehat}

\def\bra{\langle}
\def\ket{\rangle}
\def\sgn{{\rm sgn}}
\def\dd{{\rm d}}

\def\de{\delta}

\def\ve{\varepsilon}
\def\vp{\varphi}

\def\mR{\mathcal{R}}
\def\mH{\mathcal{H}}

\usepackage{color}

\begin{document}

\title{The separate universe framework in group field theory condensate cosmology}

\author{Florian Gerhardt} \email{florian.gerhardt@aei.mpg.de}
\affiliation{Max Planck Institute for Gravitational Physics (Albert Einstein Institute),\\
Am M\"uhlenberg 1, 14476 Golm, Germany, EU}

\author{Daniele Oriti} \email{daniele.oriti@aei.mpg.de}
\affiliation{Max Planck Institute for Gravitational Physics (Albert Einstein Institute),\\
Am M\"uhlenberg 1, 14476 Golm, Germany, EU}
\affiliation{II Institute for Theoretical Physics, University of Hamburg, Luruper Chaussee 149, 22761 Hamburg, Germany, EU} 

\author{Edward Wilson-Ewing} \email{edward.wilson-ewing@unb.ca}
\affiliation{Department of Mathematics and Statistics, University of New Brunswick, Fredericton, NB, Canada E3B 5A3}
\affiliation{Max Planck Institute for Gravitational Physics (Albert Einstein Institute),\\
Am M\"uhlenberg 1, 14476 Golm, Germany, EU}

\begin{abstract}

We use the separate universe framework to study cosmological perturbations within the group field theory formalism for quantum gravity, based on multi-condensate quantum states.  Working with a group field theory action for gravity minimally coupled to four scalar fields that can act as a set of relational clock and rods, we argue that these multi-condensate states correspond to cosmological space-times with small long-wavelength scalar perturbations.  Equations of motion for the cosmological perturbations are derived, which in the classical limit agree with the standard results of general relativity and also include quantum gravity corrections that become important when the space-time curvature approaches the Planck scale.

\end{abstract}

\maketitle

\section{Introduction}
\label{s.intro}

Group field theory (GFT) is a candidate formalism for quantum gravity that can be seen as a second-quantized version of loop quantum gravity and of simplicial geometry (for recent reviews see, e.g.,~\cite{Oriti:2011jm, Krajewski:2012aw, Oriti:2017ave}).  In GFT the fundamental excitations correspond to spin-network nodes or, equivalently, simplicial building blocks, that can be combined to construct the spin-network states of loop quantum gravity \cite{Oriti:2013aqa} as well as extended simplicial complexes of three topological dimensions (`discrete quantum spaces'). These building blocks can be viewed as quanta of space; each carry quantum numbers that capture information about surface areas and 3-volumes, as well as the rest of the (quantum) discrete geometries that can be associated to general ensembles of such building blocks. The discrete geometric data are the basis for constructions aiming to extract approximate continuum physics from the fundamental models.

In particular, the total volume of a `universe' associated to a given state made of $N$ quanta is obtained by adding up the volume contributions associated to each of the quanta. It seems reasonable to assume that the quantum gravity state corresponding to (a constant-time slice of) a classical space-time with a volume that is large compared to the Planck volume is highly excited, in the sense that it corresponds to a superposition of quantum states of which most are composed of a large number of quanta of geometry with each individual excitation contributing a Planck-scale volume to the total. In other words, the study of continuum physics within the full quantum gravity formalism requires control over its non-perturbative sector (in terms of the number of fundamental excitations involved), going beyond what is captured by simple spin-network states with only a few quanta of geometry.

This motivates the study of GFT condensate states, the simplest states of this non-perturbative (in excitation number) type.  Further, the symmetries of the simplest condensates (where all excited quanta are in the same state) are analogous to the homogeneity of cosmological space-times, and more generally the domain of reduced 1-particle densities is isomorphic to the minisuperspace of homogeneous geometries, thus suggesting that GFT hydrodynamics can be given a cosmological interpretation and that, in particular, GFT condensate states may correspond to cosmological space-times \cite{Gielen:2013kla, Gielen:2013naa}.  Indeed, for a GFT corresponding to gravity minimally coupled to a massless scalar field, not only do the emerging dynamical equations for the spatial volume (with the scalar field playing the role of a relational time variable) correspond exactly to the usual Friedmann equations in the classical limit, but also quantum gravity corrections become important in the Planck regime and ensure that the volume never vanishes, thereby replacing the big-bang singularity by a non-singular quantum bounce \cite{Oriti:2016qtz, Oriti:2016ueo}.  This is one key result obtained in the context of `GFT condensate cosmology', for a review see, e.g.,~\cite{Gielen:2016dss, Oriti:2016acw}.

To further connect this formalism with physical cosmology, among other tasks, it is important to understand how to include inhomogeneous perturbations.  One proposal is to allow the condensate to depend on position, with position being measured using matter fields that act as relational rods \cite{Gielen:2017eco}.  In this paper we further develop this proposal using the separate universe approach to describe long-wavelength scalar perturbations \cite{Wilson-Ewing:2015sfx}.

More specifically, in the separate universe framework a cosmological space-time is approximated by a collection of large homogeneous patches that each have the line element of the Friedmann-Lema\^itre-Robertson-Walker (FLRW) space-time, although with a scale factor and lapse that vary from one patch to another.  The long-wavelength scalar perturbations are captured in the differences in the scale factor and lapse between patches.  Here we consider a collection of GFT condensate states (combined into a single multi-condensate state), with each condensate corresponding to the portion of the space-time in one patch.

Most studies of quantum gravity effects in cosmological perturbation theory use in an essential way the preferred coordinates singled out by homogeneity and isotropy on spatial slices, either by expanding around a saddle-point solution in the path integral on which coordinates can be introduced \cite{Halliwell:1984eu, Feldbrugge:2017kzv} or through some gauge-fixing before quantization \cite{FernandezMendez:2012vi, WilsonEwing:2012bx, Agullo:2012fc}.  Instead, here a relational framework is used to provide relational coordinates with respect to matter fields.  In this way, in GFT condensate cosmology (relational) coordinates can be defined after quantization, rather than being required in order to derive the quantum equations of motion for the system.

In order to extract the equations of motion for the long-wavelength scalar perturbations from the multi-condensate state (whose dynamics are determined by the GFT action), since the matter fields will in principle evolve differently in different patches it is first necessary to determine how to define a spatial slice (or equivalently, an instant of equal time in all of the different patches). This is the main conceptual obstacle that we face in this first step beyond homogeneous cosmology via GFT condensates. Once this has been achieved, it is possible to derive the equations of motion for the cosmological perturbations. These turn out to be coupled equations of motion for a number of variables describing the perturbations, but for scalar modes (in the absence of anisotropic stress) there is only one gauge-invariant degree of freedom. The last step is to isolate the equation of motion for the single degree of freedom; it turns out that for a massless scalar field this last step is quite direct. This equation of motion has the correct semi-classical limit and reduces to the one found in loop quantum cosmology for a simple class of GFT states.

These results show how, for a certain class of quantum states in the GFT formalism for quantum gravity, it is possible to study cosmological perturbations, at least within the separate universe approximation.  The resulting equation for long-wavelength scalar cosmological perturbations can be used in future works to determine whether, for the GFT states considered here, quantum gravity effects could leave some detectable signature, e.g., in the cosmic microwave background.  We leave this question for future work.  Another important open question is to understand how to handle cosmological perturbations outside of the separate universe approximation, in particular to extend these results to short-wavelength perturbations, this is also left for future work.

\section{GFT Condensate Cosmology}
\label{s.flrw}

The idea underlying the separate universe framework is essentially to discretize long-wavelength scalar perturbations (on a flat FLRW background) on a lattice \cite{Wilson-Ewing:2015sfx}.  This lattice is constituted of a large number of patches, each approximated as being homogeneous (thus with a four-dimensional geometry that is fully captured by a scale factor and a lapse function, i.e., by the flat FLRW metric), and the inhomogeneities are encoded in the differences in the scale factors, lapse functions and energy densities of the matter field in the different patches. The dynamics of the perturbations are then determined by the evolution of each homogeneous patch, neglecting interactions between patches.

Previous work has shown how the hydrodynamics of a simple GFT condensate state gives the Friedmann equations (with some quantum gravity corrections), suggesting that these simple condensate states once coarse-grained correspond to flat FLRW space-times, and therefore are the cosmological sector of GFT \cite{Gielen:2013kla, Gielen:2013naa, Oriti:2016qtz, Oriti:2016ueo}.  Now, following the separate universe framework, we will study perturbations by constructing states that are tensor products of these condensate states, with each condensate corresponding to one `FLRW' patch in the separate universe lattice.  Thus, before setting up the separate universe framework in the GFT context, we briefly review the basic GFT condensate cosmology results, obtained for single condensate states corresponding to homogeneous universes.

\subsection{Group Field Theory}
\label{ss.gft}

The GFT model we will consider here is for Lorentzian quantum gravity minimally coupled to a massless scalar field, in a parametrised form encompassing the so-called EPRL models proposed in the context of loop quantum gravity and spin foam models \cite{Engle:2007wy}, but also general enough to include other models built on similar criteria \cite{Baratin:2010wi, Baratin:2011hp}.  These models are quantum field theories for a field over $SU(2)^{\times 4} \times \mathbb{R}$, with the Lorentz covariance of the theory encoded in the kernels defining the action (which in turn determines the dynamics for the GFT). The $SU(2)$ variables admit an interpretation in terms of a discrete gravity connection living on the simplicial structures generated by the theory, and the real variables define a discretized real scalar field living on the same lattices. Here we only use the presentation of the theory in terms of $SU(2)$ representation data, corresponding to eigenvalues of discrete geometric quantities \cite{Oriti:2013aqa}. The fundamental operators are the field operators $\h\vp^{j_i,\iota}_{m_i}(\phi)$ and $(\h\vp^{j_i,\iota}_{m_i})^\dag(\phi)$ that act as annihilation and creation operators respectively.  These operators have 4 pairs of $SU(2)$ labels $j, m$ and one $SU(2)$ intertwiner $\iota$, as well as one continuous label corresponding to the massless scalar field $\phi$.

The field operators are assumed to satisfy bosonic commutation relations,
\be
[\h\vp^{j_i,\iota}_{m_i}(\phi), (\h\vp^{\t j_i,\t \iota}_{\t m_i})^\dag(\t\phi)] =
\prod_{i=1}^4 \left( \de^{j_i, \t j_i} \de_{m_i, \t m_i} \right) \de^{\iota, \t\iota} \de(\phi - \t\phi),
\ee
and $[\h\vp, \h\vp] = 0 = [\h\vp^\dag, \h\vp^\dag]$.  The GFT Fock vacuum $|0\ket$ is the state annihilated by all $\h\vp$:
\be
\h\vp^{j_i,\iota}_{m_i} |0\ket = 0.
\ee
Finally, physical states $|\Psi\ket$ are those that satisfy the quantum equations of motion
\be \label{gft-eom}
\wh{\f{\de S}{\de \varphi}} |\Psi\ket = 0,
\ee
and its hermitian conjugate, where $S[\vp, \b\vp]$ is the GFT action.

The GFT action has the form $S = K - V$ with
\be
K = \int \dd \phi \, \dd \t\phi \sum_{j_i, \t j_i, m_i, \t m_i, \iota, \t\iota} K^{j_i, \t j_i,\iota, \t\iota}_{m_i, \t m_i}((\phi-\t\phi)^2) \, \vp^{j_i,\iota}_{m_i}(\phi) \, \b\vp^{\t j_i, \t \iota}_{\t m_i}(\t\phi),
\ee
and $V$ is composed of a sum of fifth order interaction terms in the GFT fields $\vp$ and $\b\vp$ that is local in the matter label $\phi$.  Note that in general $K^{j_i, \t j_i,\iota, \t\iota}_{m_i, \t m_i}((\phi-\t\phi)^2)$ could in principle depend on $\phi$ and $\t\phi$ in a more complicated way rather than the combination $(\phi-\t\phi)^2$, but such a theory would not capture the symmetries of gravity minimally coupled to a massless scalar field, namely $\phi \to -\phi$ and $\phi \to \phi + \phi_o$.

The parameters in the action are chosen to ensure that the action $S$ be real-valued.  The combinatorics and polynomial order of the interaction is chosen so that the Feynman diagrams of the model are simplicial complexes, while the kernels are chosen so that the corresponding Feynman amplitudes are quantum simplicial gravity path integrals for 4d Lorentzian gravity coupled to a free, real, massless scalar field \cite{Oriti:2016qtz, Oriti:2016ueo, Li:2017uao}.

\subsection{Condensate States}
\label{ss.cond}

The family of condensate states $|\sigma\ket$ considered here is built on the simplest condensate states corresponding to coherent states of the field operator:
\be \label{def-cond}
|\sigma\ket = \exp \left( \int \dd\phi \sum_{j_i, m_i, \iota} \sigma^{j_i,\iota}_{m_i}(\phi) \, (\h\vp^{j_i,\iota}_{m_i})^\dag(\phi) \right) |0\ket \quad,
\ee
where $\sigma^{j_i, \iota}_{m_i}(\phi)$ is the condensate wave function.  The property that all GFT quanta entering the definition of such states are assigned the same wave function is the quantum counterpart of the homogeneity of the corresponding continuum geometry \cite{Gielen:2016dss, Oriti:2016acw}.

Consistently with this interpretation, these quantum states, although encoding a superposition of highly excited states with respect to the GFT Fock vacuum, are fully captured by a single condensate wave function which lives on a domain of geometric data (corresponding to the phase space of geometries of a single tetrahedron) that is isomorphic to the minisuperspace of homogeneous continuum geometries \cite{Gielen:2013naa}.  This provides a direct mechanism to compute collective observables relevant to cosmology, but must be used with care. Let us discuss this point in some more detail, to clarify both its significance and its limitations. 

This isomorphism is, more generally, between the domain of the `reduced 1-particle density function' of GFT models for 4d quantum gravity (i.e., those related to simplicial gravity path integrals and spin foam models) on which any hydrodynamic approximation would be based, and the minisuperspace of homogeneous continuum geometries. This is the general basis for interpreting GFT hydrodynamics as cosmological dynamics.  A key assumption underlying GFT condensate cosmology is that the `cosmological phase' of GFT is a condensate phase; as a result calculations are more direct and there exists a closer relation between the microscopic quantum GFT dynamics, where results in simplicial quantum geometry and loop quantum gravity can be used, and the effective hydrodynamics of the system (here, cosmological dynamics) than there would otherwise be. In this case, the reduced 1-particle density becomes the condensate wave function. The further simplification of considering condensate states of the form \eqref{def-cond}, which lead to mean field Gross-Pitaevskii hydrodynamics, implies that in this case the condensate wave function coincides with the 1-particle wave function for individual tetrahedra forming the condensate state. This case is even more convenient for calculations, because the relation between quantum gravity microphysics and cosmology is truly direct. Of course, this class of states is probably too simple to be fully realistic, and the very direct correspondence between microscopic and macroscopic observables should be taken with caution: in some cases it may be approximately correct, while in others it may break down.

For example, on the one hand the total spatial volume of the space-time (at an instant of relational time) is clearly given by the sum of the volumes of each of the quanta of geometry, so (cosmological) volume dynamics, due to the combination of the change in the number of quanta of geometry and in the average volume of each individual quantum, may well be captured by the simplest condensate states.  On the other hand, one may well suspect that this Gross-Pitaevskii mean field approximation likely fails already in accounting correctly for anisotropic dynamics as this would require too strong of a correspondence between the dynamics of cosmological anisotropies and the microscopic anisotropies of the fundamental building blocks of the universe: as the universe expands, the anisotropies would be required to relax at exactly the same rate microscopically, in each individual GFT excitation, and macroscopically, in the emergent cosmological space-time. This suggests that, while it may be possible to describe the anisotropic Bianchi space-times within the context of GFT hydrodynamics or even GFT condensate hydrodynamics, it will likely be necessary to go beyond the Gross-Pitaevskii approximation considered here in \eqref{def-cond}.

\

Nonetheless, since we are interested, at least to start, in the relatively simple space-times of homogeneous and isotropic universes, it seems reasonable not only to start with states of the form \eqref{def-cond}, but also to further simplify the form of the condensate state $|\sigma\ket$ by restricting attention to a simple sub-family corresponding to those where the condensate wave function is isotropic in the sense that it only has support on configurations corresponding to isotropic quanta of geometry, identified with equilateral tetrahedra:
\be \label{sigma1}
\sigma^{j_i,\iota}_{m_i}(\phi) = \de^{\iota,\iota^\star} \left( \prod_{i=2}^4 \de_{j,j_i} \right) \bar{\mathcal{I}}^{jjjj,\iota^\star}_{m_1m_2m_3m_4} \, \sigma_j(\phi),
\ee
where $j_1 = j$ and $\iota^\star$ labels the intertwiner $\mathcal{I}$ that maximizes the expectation value of the loop quantum gravity volume operator (or of its simplicial quantum gravity counterparts) on a 4-valent spin-network node with 4 spins $j$.  Condensate states of this type are thus fully determined by a condensate wave function $\sigma_j(\phi)$ which depends only on one $j$ and on $\phi$.

States corresponding to physical configurations allowed by the GFT are those that satisfy the quantum equation of motion \eqref{gft-eom}.  In simple condensate states, interaction terms are typically subdominant so we neglect the potential term in the equations of motion as a first approximation.  In addition, to simplify the form of the kinetic term, we rewrite $\t\phi = \phi + \de\phi$, perform a Taylor expansion in the action around $\de\phi = 0$, and keep the first two terms, higher order terms being suppressed by higher powers of $\hbar$.  This is also sensible in a mean field hydrodynamic analysis as the one we are interested in performing here. The resulting quantum equation of motion for $|\sigma\ket$ gives an equation for the condensate wave function $\sigma_j(\phi)$,
\be \label{eom-sigma-flrw}
\partial_\phi^2 \sigma_j(\phi) - m_j^2 \sigma_j(\phi) = 0,
\ee
where $m_j$ depends on the parameters appearing in the GFT action, and specifically in the relative weights of the first and second terms in the Taylor expansion of the kinetic term \cite{Oriti:2016qtz, Oriti:2016ueo}.

The above constitutes the equation of motion for the chosen quantum state coming from the GFT model we work with.  However, in homogeneous and isotropic cosmology, the main quantity of interest is the spatial volume $V$ (i.e., the cube of the scale factor appearing in the metric).  The (expectation value of the) volume at an instant of relational time $\phi$ is given by
\be
V(\phi) = \bra \sigma | \h V(\phi) | \sigma \ket = \sum_j |\sigma_j(\phi)|^2 V_j,
\ee
with $V_j \sim j^{3/2} \lp^3$ being the approximate eigenvalue of the 1st quantized volume operator acting upon a spin-network node with 4 spins $j$ and the intertwiner $\iota^\star$.

Clearly, a $V(\phi)$ computed from a condensate wave function that solves \ref{eom-sigma-flrw} encodes the relational dynamics for $V$ with respect to $\phi$. This is what gives the emergent modified Friedmann equations. They show two key features. If $m_j^2 \approx 3 \pi G$, then these equations reduce to the usual Friedmann equations of general relativity in the classical limit (while sufficient, this condition is not necessary to give the correct semi-classical dynamics \cite{Gielen:2016uft}), while quantum gravity corrections generate a non-singular bounce to occur in the Planck regime which replaces the big-bang singularity of general relativity \cite{Oriti:2016qtz, Oriti:2016ueo}.

In the case that the condensate wave function $\sigma_j(\phi)$ is non-vanishing only for one value of $j=j_o$, the emergent modified Friedmann equations simplify to \cite{Oriti:2016qtz, Oriti:2016ueo}
\be
\left( \f{1}{3V(\phi)} \, \f{\dd V(\phi)}{\dd \phi} \right)^2 =\f{4 \pi G}{3} \left(1 - \f{\rho(\phi)}{\rho_c}\right) - \f{4 V_{j_o} E_{j_o}}{9 V(\phi)},
\ee
\be
\f{1}{V(\phi)} \, \f{\dd^2 V(\phi)}{\dd^2 \phi} = 12 \pi G - \f{2 V_{j_o} E_{j_o}}{V(\phi)},
\ee
where $\rho(\phi) = \pi_\phi^2 / 2 V(\phi)^2$ is the matter energy density, with $\pi_\phi$ (the momentum of the scalar field) a constant of the motion, $\rho_c = 3 \pi G \hbar^2 / 2 V_{j_o}^2 \approx (3 \pi / 2 j_o^3) \rho_{\rm Pl}$ is the critical energy density, and $E_{j_o}$ is a state-dependent constant of the motion.  

Note that if $E_{j_o}=0$, these reduce to the LQC effective equations expressed in terms of the relational clock $\phi$ \cite{Ashtekar:2006wn}.

\section{Separate Universes in the GFT formalism}
\label{s.sep}

The separate universe framework was first motivated by the observation that the spatial derivative term in the equations of motion for cosmological perturbations becomes negligible for Fourier modes whose wavelength is greater than the Hubble radius (or the sound radius if the sound speed is not equal to one) \cite{Salopek:1990jq, Wands:2000dp}.  In this sense, at large (super-Hubble) scales cosmological perturbations at different locations can be understood to evolve independently: interactions between perturbations at different locations can safely be neglected.

In fact, this observation can be used be derive the dynamics for long-wavelength scalar perturbations directly from the Friedmann equations of a flat FLRW space-time, thus using only the dynamics of homogeneous universes \cite{Wilson-Ewing:2015sfx}.  To see this, recall that the line element of the spatially flat FLRW space-time with scalar perturbations, expressed in the longitudinal gauge and conformal time, is
\be \label{ds2}
\dd s^2 = - a(t)^2 \Big( 1 + 2 \psi(\vec x, t) \Big) \dd t^2 + a(t)^2 \Big( 1 - 2 \psi(\vec x, t) \Big) \dd \vec x^2,
\ee
assuming the matter content has vanishing anisotropic stress.

In general relativity (or in the limit where general relativity provides a good approximation to the underlying quantum gravity theory), for the case of the matter content being a scalar field $\phi$, the co-moving curvature perturbation is
\be
\mR = \f{\mH}{\phi'} \de\phi + \psi,
\ee
with $\de\phi$ the perturbation in the matter field, primes denoting derivatives with respect to conformal time and $\mH = a'/a$ the conformal Hubble rate.  The power spectrum $\Delta_\mR^2(k)$ of $\mR$ can be expressed in terms of its Fourier modes,
\be
\Delta_\mR^2(k) = \f{k^3}{2 \pi^2} |\mR_k|^2,
\ee
and is usually parametrized by $\Delta_\mR^2(k) = A \, k^{n_s-1}$; observations of the cosmic microwave background indicate that $A \sim 10^{-9}$ and $n_s = 0.968 \pm 0.006$ \cite{Ade:2015xua}.  For more on cosmological perturbation theory in general relativity see, e.g.,~\cite{Mukhanov:1990me}.  While the observed near-scale-invariance can be explained by inflation (or alternatives like ekpyrosis or the matter bounce), it is possible that there could be some sub-leading effects due to quantum gravity present. It is also possible that quantum gravity will provide an altogether new mechanism for producing near-scale invariance in the cosmological power spectrum.  For this reason, it is important to understand how quantum gravity effects could modify the dynamics of cosmological perturbations and, more generally, to develop a theory of cosmological perturbations within a fundamental quantum gravity formalism.

Returning to the separate universe framework, if the metric \eqref{ds2} is discretized over $n_{tot}$ (super-Hubble) patches, each approximately homogeneous, then the line element in each patch $n$ is
\be
\dd s^2 = - N_n(t)^2 \dd t^2 + a_n(t)^2 \dd \vec x^2,
\ee
with $N_n(t) = a(t) [1 + \psi_n(t)]$ and $a_n(t) = a(t) [1 - \psi_n(t)]$, where $\psi_n(t)$ is the average value of $\psi(\vec x, t)$ in the patch $n$ at time $t$.  This is precisely the line element for a flat FLRW space-time, with a lapse and scale factor whose values vary from patch to patch.  Then, since long-wavelength perturbations evolve independently, the dynamics of each $a_n(t)$ is determined by the Friedmann equation, for the choice of the lapse $N_n(t)$.  In general relativity, the resulting equations of motion are precisely those for long-wavelength scalar perturbations in the longitudinal gauge.

In modified gravity theories (for example, theories that include corrections coming from quantum gravity effects) where the dynamics for FLRW space-times are known, the separate universe framework can be used to derive the equations of motion for long-wavelength scalar perturbations, without relying on a complete treatment of inhomogeneous geometries.  For example, this has been done successfully in loop quantum cosmology \cite{Wilson-Ewing:2015sfx, WilsonEwing:2012bx}.

The aim of this paper is to use the separate universe framework to derive the equations of motion for long-wavelength scalar perturbations in GFT condensate cosmology.  The basic idea is to construct a condensate state wherein there are many patches, with a condensate wave function in each patch, and thus each governed by the modified Friedmann equations given above in Sec.~\ref{ss.cond}.  For such a state, there would exist collective observables in each patch like the expectation value for the total volume $V_n$ of a given patch at an instant of relational time, also defined differently in each patch.  This particular collective observable will play a central role in the following, since it can be directly related to the scale factor $a$ and perturbation $\psi_n$ by $V_n = a^3 (1 - 3 \psi_n)$.

However, some work is required in order to make this identification precise.  Specifically, in order to include cosmological perturbations of any type there must be a way to localize excitations in space as well as in time.  A matter field has been used successfully as a relational clock, and this suggests introducing three additional matter fields to use as relational rods that will provide a relational (and physical) coordinate system \cite{Gielen:2017eco}.  Following \cite{Brown:1994py}, we introduce three dust fields that can act as a relational Cartesian coordinate system.  An advantage of using dust fields, given that a massless scalar field is already present, is that their effect on the dynamics of the space-time is expected to be negligible at high curvatures in comparison to the scalar field and therefore can safely be neglected, hence simplifying the calculations.

We now construct the GFT condensate state of interest, using some approximations suitable to the separate universe framework, and explain how to use these relational rods and clock to provide an explicit map from the observables in the GFT condensate state to the standard variables of cosmology in Secs.~\ref{ss.rel} and Sec.~\ref{ss.synch}.  Then in Sec.~\ref{ss.perts} we determine how the approximate dynamics of the GFT condensate state, determined by the GFT action, translate into equations of motion for the volume in each patch and hence for $a$ and $\psi_n$. In the classical limit these equations of motion agree exactly with those of general relativity (and in semi-classical effective field theory) and also include quantum gravity corrections that are important when the space-time curvature is of the order of the Planck scale.  We end by considering two limiting cases of interest where the equations of motion simplify considerably.

\subsection{The Condensate State and Relational Observables}
\label{ss.rel}

The GFT state used for a separate universe approach to cosmological perturbations shall be a product of many condensate states, with each condensate corresponding to a homogeneous patch.  A key point here is the central assumption in the separate universe approximation that the condensate state in each patch evolves independently of the states in other patches.

In order to define such state (and also point out the approximations we are going to use), we can start with the simple Gross-Pitaevskii GFT condensate states introduced in \cite{Gielen:2017eco}, including four scalar fields, one used as a clock ($\phi$) and three as rods ($r_i$):
\be \label{inhomowave}
| \Psi \ket = \exp\left( \int \dd\phi\, \int \dd^3 r_i \, \hat{\sigma}(r_i,\phi)\right) \; | 0\ket 
\ee
where the shorthand
\be \label{short}
\hat{\sigma}(r_i,\phi)=\sum_{j_i, m_i, \iota} \sigma^{j_i,\iota}_{m_i}(\phi, r_i) \, (\h\vp^{j_i,\iota}_{m_i})^\dag(\phi,r_i)
\ee
is used in order to focus on the dependence of the condensate wave function on the relational clock and rods. We assume the domain of the three $r_i$ scalar fields to be $\mathbb{T}^3$, and we identify the surfaces $r_i = 0$ and $r_i = 1$ (for each $i$ separately).  The condensate wave function we shall consider here is a straightforward generalization of \eqref{sigma1}, now with a dependence on the relational rods $r_i$ as well.

The approximation used in \cite{Gielen:2017eco} to study cosmological perturbations around a homogeneous universe, starting from the state \eqref{inhomowave} is to study perturbations in the GFT condensate wave function: $\sigma(r_i,\phi) \, \approx \, \sigma_0(\phi)\, +\, \psi(r_i,\phi)$.

We now want to approximate this state in a different way motivated by the separate universe framework, namely by rewriting \eqref{inhomowave} as a product state, with each component associated to a single patch.  A straightforward way to do this is to discretize the domain of the scalar field rods, which localize the quantum geometric data in space, in a cubic lattice.  For simplicity, we assume that each patch corresponds to a (spatial) cube with a non-vanishing (relational coordinate) volume.  The homogeneous patches forming the separate universe space-time can then each be defined by intervals in the three $r_i$, i.e., for each patch $n$ the relational rods satisfy $r_i \in [(r_i)_{min}^n, (r_i)_{max}^n]$.  For simplicity, we choose the $(r_i)_{min}^n$ and $(r_i)_{max}^n$ so that the relational coordinate lengths of each side of all cubes are equal (and therefore so are the volumes of each cubic patch).

Following the discussion above, note that $\int \dd^3 r_i \, f(r_i) = \sum_n \dd^3 r_i \, W_n(r_i) f(r_i)$ for any $f(r_i)$ where the window function $W_n(r_i)=1/W_o$ for $r_i \in [(r_i)_{min}^n, (r_i)_{max}^n]$ and vanishes elsewhere (the constant $W_o$ is chosen so that $\int \dd^3 r_i W_n(r_i) = 1$, the same $W_o$ can be used in each patch since they all have the same volume with respect to the $r_i$ relational coordinates).

Using this identity, the state \eqref{inhomowave} can be rewritten as
\begin{eqnarray}
|\Psi\ket &=& \exp\left[ \int \dd\phi \,\left( \sum_n \,\int \dd^3 r_i \, W_n(r_i) \hat{\sigma}(\phi, r_i) \right)\right]\, |0\ket \nn \\
&=&  \exp\left[ \int \dd\phi \,\left( \sum_n \, \hat{\sigma}_n(\phi) \right)\right]\, |0\ket  \nn \\
&=& \prod_n \, \exp \left[ \int \dd\phi \,\, \hat{\sigma}_n(\phi) \right]\, |0\ket,  \label{separatewave}
\end{eqnarray}
where
\be
\hat{\sigma}_n(\phi) = \int \dd^3 r_i \, W_n(r_i) \hat{\sigma}(\phi, r_i).
\ee
Clearly, in this way the state $|\Psi\ket$ can be seen as a product of condensate states, one in each patch.  So far, this is exact.  Now, we shall make two approximations motivated by the separate universe framework.  First, we assume that the condensate wave function $\sigma_j(r_i,\phi)$ is constant with respect to the $r_i$ in each patch $n$, to make this explicit we write the condensate wave function in the patch $n$ as $\sigma_{n,j}(\phi)$.  Second, we assume that interactions between patches are negligible and therefore the quantum equations of motion in each patch are identical to those for a single condensate reviewed in Sec.~\ref{ss.cond}, and therefore each $\sigma_{n,j}(\phi)$ is assumed to satisfy the equation of motion \eqref{eom-sigma-flrw}.  As a result, the equations of motion in this approximation are independent of the $r_i$.

A number of observables can be computed for $|\Psi\ket$, with the most interesting for our purpose being coarse-grained relational observables in each patch that can be used to extract equations of motion for the emergent cosmological space-time.  One key observable is the spatial volume of a patch at an `instant' of relational time $\phi$,
\begin{align} \label{vol-op}
V_n(\phi) = \bra \Psi | \h V_n(\phi) | \Psi \ket & =
\bra \Psi | \int \dd^3 r_i \, W_n(r_i) \sum_{j_i,m_i,\iota_v} V_{j_i,\iota_v} \vp^{j_i,\iota_v}_{m_i}(\phi, r_i)^\dag \vp^{j_i,\iota_v}_{m_i}(\phi, r_i) | \Psi \ket \nn \\
& = \sum_j V_{j,\iota^\star} |\sigma_{n,j}(\phi)|^2,
\end{align}
where, as in the homogeneous case, the $\iota_v$ label the intertwiners that are eigenstates of the loop quantum gravity volume operator (or of the analogous volume operator in quantized simplicial geometry) and $V_{j_i,\iota_v}$ denotes their eigenvalue.  Note that the above result follows from the normalization of the intertwiners, $\sum_{m_i} \bar{\mathcal{I}}^{jjjj,\iota^\star}_{m_1m_2m_3m_4} \mathcal{I}^{jjjj,\iota^\star}_{m_1m_2m_3m_4} = 1$. It should be obvious that $V_{tot}(\phi) = \bra \Psi | \h V_{tot}(\phi) | \Psi \ket =  \bra \Psi | \sum_n \h V_n(\phi) | \Psi \ket = \sum_n V_n(\phi)$. 

Another important observable for cosmology is the momentum of the massless scalar field $\phi$, which is constructed in a similar fashion in each patch as:
\begin{align} \label{pi-op}
(\pi_\phi)_n(\phi) & = 
\f{\hbar}{2i} \bra \Psi | \int \dd^3 r_i \, W_n(r_i) \sum_j \left[ \vp_j(\phi, r_i)^\dag \partial_\phi\vp_j(\phi, r_i)
- \partial_\phi \vp_j(\phi, r_i)^\dag \vp_j(\phi, r_i) \right] | \Psi \ket \nn \\
& = \f{\hbar}{2i} \sum_j \Big[ \b \sigma_{n,j}(\phi) \partial_\phi \sigma_{n,j}(\phi) - \sigma_{n,j}(\phi) \partial_\phi \b \sigma_{n,j}(\phi) \Big].
\end{align}

These are relational observables, in each patch, with respect to the massless scalar field that is acting as a clock.  To compare observables in different patches at some instant of time requires a notion of simultaneity, i.e., some way to synchronize the relational clocks in different patches.  This is the issue we turn to now.

\subsection{Synchronization and Time Evolution}
\label{ss.synch}

In order to be able to compare quantities at an instant of time (e.g., the volume in different patches at `equal time'), it is necessary to define an instant of simultaneity.  While $\phi$ acts as a clock variable in every patch, and this role can be formalized by the expectation value of the corresponding observable in each patch condensate state, these relational clocks are not necessarily synchronized. Indeed, in the presence of non-vanishing perturbations, for a general coordinate choice the relational clocks will not be synchronized even in the classical case.  In the GFT state $|\Psi\ket$, the lack of synchronization can be seen in the different condensate wave functions associated to the different patches.

Moreover, the usual theory of cosmological perturbations, which we aim to reproduce and generalize in the GFT quantum gravity setting, is not formulated in terms of relational clock variables, but rather in terms of coordinate time variables. Therefore, in order to make contact with the standard results of cosmological perturbation theory in general relativity, it is necessary to relate the relational clock variable $\phi$ to a coordinate time variable (or potentially even several different coordinate time variables, e.g., proper and conformal time).  This requires introducing a notion of simultaneity that goes beyond%
\footnote{To be clear, setting equal time slices to correspond to the relational instant where the relational scalar field has the same value in all patches is a perfectly good way to define simultaneity, this corresponds to the co-moving gauge in standard cosmological perturbation theory.  However, it is not sufficiently general to capture all gauges.  For example, the longitudinal gauge that is of interest in this paper, see \eqref{ds2}, is not in the co-moving gauge and therefore a more general notion of simultaneity is needed.}
`equal $\phi$ in all patches.'

This can be done by introducing a coordinate time $t$ as follows. In each cell, define the one-to-one map
\be \label{def-time}
T_n(\phi) = t,
\ee
then `equal times' in different cells $n$ correspond to the respective values of $\phi$ in each patch such that the coordinate $t$ is the same in all cells.  Since $T_n$ is one-to-one, this map defines the functions $\phi_n(t)$ and its inverse $t(\phi_n)$ in each cell%
\footnote{While $\h\phi$ is an operator in GFT, as explained above the key observables we are interested in, $V_n$ and $(\pi_\phi)_n$, can both be calculated directly from the condensate wave function $\sigma_{n,j}(\phi)$ which is simply a function that depends on $\phi$.  Therefore, for the observables of interest evaluated for the states considered here, this map is easy to implement; more work may be required to define it for more general states.}.
Then, $V_n(t) = V_n(\phi_n(t))$ and it is now possible to compare $V_n(t)$ and $V_m(t)$ in a meaningful way. 

From the function $\phi_n(t)$ obtained from the map $T_n$ and using the cosmological observables $(\pi_\phi)_n$ and $V_n$, it is possible to define a positive-definite function $N_n$ in each cell through
\be \label{def-N}
N_n = \f{V_n \, \phi_n'}{(\pi_\phi)_n}, \qquad\quad
{\rm with} \qquad \phi_n' = \f{\dd \phi_n(t)}{\dd t}.
\ee
The prefactors are chosen for later convenience so that $N_n$ will correspond with the lapse function in general relativity.  (Of course, it is also possible to define other combinations of these quantities, but they are not as useful in making contact with the results of general relativity.)  Equivalently, a positive-definite $N_n$ defines a choice for $t$ by
\be \label{free-N}
\dd t = \f{V_n}{N_n \, (\pi_\phi)_n} \, \dd\phi_n.
\ee
One can understand this procedure as re-introducing the coordinate redundancies of canonical general relativity (limited to time evolution) in our fully background independent (and coordinate-free) quantum gravity formalism, for easier comparison with the usual treatments in general relativity where coordinates are explicitly used.  Clearly there is a lot of freedom here, and indeed this is how the freedom of reparametrizing the time coordinate reappears in the relational framework.

While many choices are possible, some may be more convenient than others.  In particular, it is possible to match the lapse in the line element \eqref{ds2} by making the choice $N_n(t) = a(t)[1 + \psi_n(t)]$.  In this way, the macroscopic observables in the GFT state, $V_n(t)$, and the choice of the time function (through the choice of the lapse) coming from using the scalar field as a relational clock for the GFT state, can each be related to the terms appearing in the line element for a spatially flat FLRW space-time with scalar perturbations.  In this way, it will be possible to make contact with the usual results (expressed in the longitudinal gauge) obtained in general relativity.

\subsection{Cosmological Perturbations}
\label{ss.perts}

With the above kinematical considerations now addressed, the next step is to extract the effective dynamics for long-wavelength scalar cosmological perturbations following the separate universe approach.

The condensate state must satisfy the GFT quantum equations of motion, namely \eqref{gft-eom} and its hermitian conjugate.  The condensate states considered here are entirely determined by their condensate wave function, and therefore the \eqref{gft-eom} reduces to an equation of motion to be satisfied by the condensate wave function $\sigma_j(r_i,\phi)$.  In general, this equation of motion will depend on how $\sigma_j(r_i, \phi)$ varies with $j$, $\phi$ and the three $r_i$ (see \cite{Gielen:2017eco} for an example for a particular GFT action).

However, for the case of interest here, the equations of motion for the condensate wave function simplify considerably.  First, for the EPRL spin foam model, assuming isotropic configurations for the quanta of geometry as in \eqref{sigma1}, different spins $j$ do not interact.  Therefore, the equations of motion for each spin $j$ decouple from each other.  This is a result of the form of the kinetic and potential terms in the GFT action for the EPRL model, and is not an approximation (unlike the second simplification).  Second, under the separate universe approximation, assuming that each patch is at least as large as the Hubble radius and that physics outside the Hubble radius cannot affect local dynamics, the condensate wave function in each patch $n$ evolves independently from the other patches.  This corresponds to neglecting derivatives with respect to $r_i$ in the equations of motion.  As a result, the equations of motion are, for each patch $n$ and for each spin $j$, a differential equation for $\sigma_{n,j}(\phi)$ with respect to $\phi$ only.

Finally, as for the homogeneous case, we also assume the GFT interactions to be subdominant within each patch. As a result, the dynamics of each $\sigma_{n,j}(\phi)$ are identical to the dynamics of the single condensate wave function $\sigma_j(\phi)$ used in homogeneous cosmology given in \eqref{eom-sigma-flrw},
\be
\partial_\phi^2 \sigma_{n,j}(\phi) - m_j^2 \sigma_{n,j}(\phi) = 0.
\ee
Note that we have also assumed that any dependence of the dynamics on the relational dust fields $r_i$ is also negligible.  This last approximation is based on the fact that the contribution of a dust field to the classical dynamics (specifically, to the Friedmann equation) can safely be neglected compared to that of a massless scalar field in the early universe (the regime of interest here), and we assume this continues to be the case in the presence of quantum gravity effects.

Rewriting the condensate wave function as
\be \label{sigma}
\sigma_{n,j}(\phi_n) = \rho_{n,j}(\phi_n) e^{i \theta_{n,j}(\phi_n)},
\ee
with $\rho, \theta \in \mathbb{R}$, the equations of motion are
\be \label{eom-rho}
\f{d^2 \rho_{n,j}}{d\phi_n^2} - \f{Q_{n,j}^2}{\rho_{n,j}^3} - m_j^2 \rho_{n,j} \approx 0,
\ee
with two constants of motion in each patch,
\be
Q_{n,j} \approx \rho_{n,j}^2 \, \f{d \theta_{n,j}}{d\phi_n},
\ee
\be
E_{n,j} \approx \left( \f{d\rho_{n,j}}{d\phi_n} \right)^2 + \rho_{n,j}^2 \left( \f{d\theta_{n,j}}{d\phi_n} \right)^2 - m_j^2 \rho_{n,j}^2 .
\ee
Again, the equations of motion for each patch are identical to the equations of motion derived for FLRW space-times in GFT condensate cosmology \cite{Oriti:2016qtz, Oriti:2016ueo}.

In terms of the time variable $t$, denoting $f' = \dd f/\dd t$ and using the chain rule $df/d\phi_n = f'/\phi_n'$, \eqref{eom-rho} becomes
\be \label{eom-rho-t}
\rho_{n,j}'' - \f{\rho_{n,j}' \, \phi_n''}{\phi_n'} - \f{Q_{n,j}^2 (\phi_n')^2}{\rho_{n,j}^3} - m_j^2 (\phi_n')^2 \rho_{n,j} = 0.
\ee
Note that the conserved quantity $E_{n,j}$ can be rewritten as
\be \label{E-t}
E_{n,j} = \left( \f{\rho_{n,j}'}{\phi_n'} \right)^2 + \f{Q_{n,j}^2}{\rho_{n,j}^2} - m_j^2 \rho_{n,j}^2.
\ee

It is now possible to derive the dynamics of perturbations, encoded in the patch-specific deviations from the values of geometric and matter observables averaged over all patches.  To do this, first calculate the average values of each $Q_j$ and $E_j$ across all patches.  Assuming there are $n_{tot}$ patches,
\be
\bar Q_j = \f{1}{n_{tot}} \sum_n Q_{n,j}, \qquad
\bar E_j = \f{1}{n_{tot}} \sum_n E_{n,j},
\ee
and the conserved quantities in each patch can be re-expressed as
\be
Q_{n,j} = \bar Q_j + \delta Q_{n,j}, \qquad
E_{n,j} = \bar E_j + \delta E_{n,j}.
\ee
Using the time function \eqref{def-time}, it is also possible to define average values of the components of the condensate wave function at an instant of time $t$:
\be \label{def-b-rho}
\bar \rho_j(t) = \f{1}{n_{tot}} \sum_n \rho_{n,j}(t),
\ee
\be
\bar \theta_j(t) = \f{1}{n_{tot}} \sum_n \theta_{n,j}(t),
\ee
and then it is possible to define $\delta \rho_{n,j}(t)$ and $\delta \theta_{n,j}(t)$ through
\be
\rho_{n,j}(t) = \bar \rho_j(t) + \delta\rho_{n,j}(t),
\ee
\be
\theta_{n,j}(t) = \bar \theta_j(t) + \delta\theta_{n,j}(t).
\ee

For cosmological perturbation theory, we are interested in small departures from homogeneity.  In this context, small inhomogeneities correspond to (i) requiring that the condensate wave function is approximately the same in each patch, and (ii) the map $T_n$ is approximately the same in each patch.  More precisely, this requires that for all patches $n$ and all values of $t$
\be
|\delta\rho_{n,j}(t)| \ll \bar\rho_j(t), \qquad
|\delta\theta_{n,j}(t)| \ll |\bar\theta_j(t)|, \qquad
|\delta N_n(t)| \ll \bar N(t),
\ee
with $\b N(t) = \sum_n N_n(t) / n_{tot}$ and $\de N_n(t) = \b N(t) - N_n(t)$ (and $N_n(t)$ defined by \eqref{def-N} in each patch), together with the further requirements that, for all $n$, $|\delta E_{n,j}| \ll |\bar E_j|$ and $|\delta Q_{n,j}| \ll |\bar Q_j|$.  If these conditions are satisfied, then the GFT state $|\Psi\ket$ can be viewed as a collection of separate universes approximating small long-wavelength scalar perturbations.

The evolution equations for the above quantities follow directly from \eqref{eom-rho-t}.  Averaging the equation of motion for each patch over all patches gives the background equation of motion,
\be \label{eom-rho-bg}
\bar \rho_j'' - \f{\bar\phi'' \bar\rho_j'}{\bar\phi'} - \f{\bar Q_j^2}{\bar\rho_j^3} \left(\bar\phi'\right)^2
- m_j^2 \bar\rho_j \left(\bar\phi'\right)^2
\approx 0,
\ee
and the remaining contribution to the equation of motion \eqref{eom-rho-t} in each patch determines the dynamics for the perturbations (to linear order assuming small perturbations):
\begin{align} \label{eom-rho-pert}
\delta \rho_{n,j}'' + \f{\bar\phi'' \bar\rho_j'}{(\bar\phi')^2} \delta\phi_n'
- \f{\bar\rho_j'}{\bar\phi'} \delta\phi_n'' - \f{\bar\phi''}{\bar\phi'} \delta\rho_{n,j}''
- \f{2 \bar Q_j^2 \bar\phi'}{\bar\rho_j^3} \delta\phi_n'
+ \f{3 \bar Q_j^2 (\bar\phi')^2}{\bar\rho_j^4} \delta\rho_{n,j} & \nn \\
- \, \f{2 \bar Q_j (\bar\phi')^2}{\bar\rho_j^3} \delta Q_{n,j}
- m_j^2 (\bar\phi')^2 \delta\rho_{n,j} - 2 m_j^2 \bar\rho_j \bar\phi' \delta\phi_n'
& \approx 0,
\end{align}

With the condensate wave function in the form \eqref{sigma}, the cosmological observables defined above in \eqref{vol-op} and \eqref{pi-op} become
\be
V_n(t) = \sum_j V_j \rho_{n,j}(t)^2, \qquad (\pi_\phi)_n = \sum_j \hbar Q_{n,j},
\ee
and it is straightforward to define total values for these observables simply by summing over all patches:
\be \label{tot}
V_{tot}(t) = \sum_n V_n(t), \qquad (\pi_\phi)_{tot} = \sum_n (\pi_\phi)_n.
\ee
Note that the $(\pi_\phi)_n = \sum_j \hbar Q_{n,j}$ (and therefore $(\pi_\phi)_{tot}$ also) are constants of the motion.  

The equations of motion for $V(t)$, then, emerge from the microscopic equations of motion for $\bar\rho_j$ and $\delta\rho_{n,j}$.  These equations can be compared to the Friedmann equations of general relativity, and simiarly the dynamics of perturbations of $V_n(t)$ from the average value in each patch can be compared with the equations of motion for cosmological perturbations found in general relativity.  Note that the continuity equation in general relativity for a massless scalar field, namely that $\pi_\phi$ is a constant of the motion, is recovered in each patch already at this point.

Following the same procedure as above, $V_n(t)$ can be expressed in terms of an average value and a perturbation: $V_n(t) = \b V(t) + \de V_n(t)$, with
\be
\b V(t) = \f{V_{tot}(t)}{n_{tot}} = \sum_j V_j \b\rho_j(t)^2,
\ee
where $\b\rho_j(t)$ is defined in \eqref{def-b-rho}, and, to linear order in perturbations (i.e., dropping the $\de\rho^2$ term) one finds
\be
\de V_n(t) = 2 \sum_j V_j \b\rho_j(t) \de\rho_{n,j}(t).
\ee

Of course, just as the lapse $N_n$ variable can be averaged with $\b N(t) = \sum_n N_n(t) / n_{tot}$, with departures in each cell from the average value given by $\delta N_n = \b N(t) - N_n(t)$ as already explained above, so can the relation $\phi_n(t)$ defined in each patch:
\be
\b\phi(t) = \f{1}{n_{tot}} \sum_n \phi_n(t), \qquad
\phi_n(t) = \bar \phi(t) + \delta \phi_n(t).
\ee

Finally, to make contact with the line element \eqref{ds2}, we choose
\be
\b N(t) = a(t), \qquad \de N_n(t) = a(t) \psi_n(t).
\ee
In this way, a direct comparison will be possible between the dynamics for cosmological perturbations emerging from the GFT condensate states, and the standard equations of motion for cosmological perturbation theory expressed in the longitudinal gauge.  Of course, other choices for the time coordinate here are possible, but then the comparison to general relativity will be more difficult.

\

To extract cosmological dynamics from the GFT condensate, it is necessary to determine how the cosmological observables evolve, as determined by the dynamics of the quantum state $|\Psi\ket$.  One useful relation is
\be \label{cont}
\f{\dd}{\dd t} \left( \f{V_n(t)}{N_n(t)} \f{\dd\phi_n}{\dd t} \right) = 0,
\ee
which follows from the definition \eqref{def-N} of $N_n$ and the result that $(\pi_\phi)_n = \sum_j \hbar Q_{n,j}$ is a constant of the motion as pointed out below \eqref{tot}.  Splitting this equation into its background part and its perturbative part, the background gives the usual continuity equation for a massless scalar field in a flat FLRW space-time,
\be
\bar \phi'' + 2 \mH \bar\phi' = 0,
\ee
where primes denote derivatives with respect to $t$ with $N_n = a(1+\psi_n)$ and $\mH = a'/a$, and at linear order in perturbation theory \eqref{cont} becomes
\be \label{pert-phi}
\delta\phi_n'' + 2 \mH \delta\phi_n' - 4 \bar\phi' \psi_n' = 0,
\ee
again the standard equation for long-wavelength linear perturbations of a massless scalar field in a flat FLRW background.

In addition, \eqref{def-N} gives $\bar\phi' = \bar\pi_\phi / a^2$ for the background and
\be \label{dphi'}
\delta\phi_n' = \bar\phi' \Big[ 4 \psi_n + (\delta \pi_\phi)_n \Big]
\ee
to first order in perturbations.  Of course, since $(\pi_\phi)_n$ is a constant of the motion, so are $\bar\pi_\phi$ and $(\delta \pi_\phi)_n$, precisely as expected for a massless scalar field.

\

The equations of motion for the geometric sector follow from
\be
V_n' = 2 \sum_j V_j \rho_{n,j} \rho_{n,j}', \qquad
V_n'' = 2 \sum_j V_j \Big[ \rho_{n,j} \rho_{n,j}'' + (\rho_{n,j}')^2 \Big].
\ee
Using \eqref{eom-rho-t} and \eqref{E-t}, these equations become, respectively,
\be
V_n' = 2 \sum_j V_j \rho_{n,j} \phi_n' \sgn(\rho_{n,j}') \sqrt{E_{n,j} + m_j^2 \rho_{n,j}^2 - \f{Q_{n,j}^2}{\rho_{n,j}^2}},
\ee
and
\be
V_n'' = \f{V_n' \phi_n''}{\phi_n'} + 2 (\phi')^2 \sum_j V_j (E_{n,j} + 2 m_j^2 \rho_{n,j}^2).
\ee

As done for the matter sector, each of these equations can be split into a background equation of motion and a separate equation of motion for linear perturbations.  The first equation, recalling $\b V = a^3$, gives
\be \label{fried-bg}
\b V' = 3 a^2 a' = 2 \sum_j V_j \b\rho_j \b\phi' \sgn(\b\rho_j') \sqrt{\b E_j + m_j^2 \b \rho_j^2 - \f{\b Q_j^2}{\b\rho_j^2}},
\ee
for the background, with the conformal Hubble rate $\mH = a'/a$ given by $\b V'/3 \b V$. Then, using $\de V_n' = -9 a^2 a' \psi_n - 3 a^3 \psi_n'$, the equation of motion for the perturbations is
\begin{align} \label{fried-pert}
\psi_n' + 3 \mH \psi_n + \mH \de\phi_n' = - \f{2 \b\phi'}{3a^3} & \sum_j \, V_j \, \sgn(\b\rho_j') \Bigg[ \sqrt{ \b E_j - (\b Q_j / \b\rho_j)^2 + m_j^2 \b\rho_j^2 \,} \,\, \de\rho_{n,j} \nn \\ &
+ \f{\f{1}{2} \, \de E_{n,j} - (\b Q_j / \b\rho_j^2) \, \de Q_{n,j} + \Big( \b Q_j^2/\b\rho_j^3 + m_j^2 \b\rho_{n,j} \Big) \de\rho_{n,j}}{\sqrt{ \b E_j - (\b Q_j / \b\rho_j)^2 + m_j^2 \b\rho_j^2 \,}} \, \Bigg].
\end{align}
The second equation gives
\be \label{ray-bg}
\f{a''}{a} + 4 \mH^2 = \f{2 (\b\phi')^2}{3 a^3} \sum_j V_j (\b E_j + 2 m_j^2 \b\rho_j^2)
= \f{2 \b\pi_\phi^2}{3 a^7} \sum_j V_j (\b E_j + 2 m_j^2 \b\rho_j^2)
\ee
for the background variables, and
\be \label{ray-pert}
\psi_n'' + 12 \mH \psi_n' = - \f{2 \b\pi_\phi^2}{3 a^7} \sum_j V_j \Big( (\b E_j + 2 m_j^2 \b\rho_j^2) [11 \psi_n + 2(\de\pi_\phi)_n]
+ \de E_{n,j} + 4 m_j^2 \b\rho_j \de\rho_{n,j} \Big),
\ee
for the cosmological perturbations, using \eqref{dphi'} to substitute out the $\de\phi_n'$ term, and the background equation \eqref{ray-bg} to simplify the expression.

These are the equations of motion for the background and linear long-wavelength scalar perturbations.  They clearly depend on the underlying quantum gravity state $|\Psi\ket$ through $\rho_{n,j}, E_{n,j}$ and $Q_{n,j}$.

\

In the classical limit, $\rho_{n,j}$ is large and in this limit it is straightforward to check that the usual dynamics of general relativity are recovered, for example,  for the choice of $m_j^2 = 3 \pi G$ for the parameter appearing in the GFT action.  The latter are sufficient conditions, not necessary ones, and the classical equations of motions can be recovered more generally.  For example, another possibility that gives the correct (general relativistic) classical limit of the dynamics is the case of only one spin $j_o$ (or only a few spins) being excited in the condensate state and $m_{j_o}^2 = 3 \pi G$.  Note that such a regime is automatically reached in the late cosmological evolution for some natural choices of the GFT action \cite{Gielen:2016uft, Pithis:2016wzf, Pithis:2016cxg}.

Away from the classical limit, the equations we have obtained for the cosmological perturbations are rather complicated, and depend explicitly on the parameters in the condensate wave function, as should be expected.  It would be nice to be able to rewrite the equations (as much as possible) in terms of cosmological variables only, but this does not appear possible for generic choices for the condensate wave function and the GFT action.  However, in some cases simplifications do arise, as shall be shown below.

In the general case where no further simplifications arise, the emergent cosmological dynamics (background and linear long-wavelength scalar perturbations) arising from a GFT condensate state $|\Psi\ket$ can perhaps most easily be determined by: (i) solving the dynamics of the condensate wave function given in \eqref{eom-rho-bg} and \eqref{eom-rho-pert} to find $\b\rho_j(t)$ and $\de\rho_{n,j}(t)$; then (ii) using the relations $a^3 = \sum_j V_j \b\rho_j^2$ and $-3 a^3 \psi_n = 2 \sum_j V_j \b\rho_j \de \rho_{n,j}$ to determine $a(t)$ and $\psi_n(t)$.  Note that in this general case it may not be possible to give initial conditions only in terms of $a(t_o)$ and $\psi_n(t_o)$ and their first derivatives, as in general relativity; rather, initial conditions concerning the full GFT state may be required, i.e., $\b\rho_j(t_o)$ and $\de\rho_{n,j}(t_o)$ and their first (relational) derivatives, together with all of $\b E_j, \b Q_j, \de E_{n,j}$ and $\de Q_{n,j}$.

\subsection{The Equal $m_j$ Case}
\label{ss.mj}

One case in which the equations of motion simplify is when all the $m_j$, the parameters appearing in the GFT action, are equal: $m_j = m$.  In this case, in order to have the correct classical limit, $m^2 = 3 \pi G$.

Even in this case the equations \eqref{fried-bg} and \eqref{fried-pert} do not simplify significantly, but the other two equations \eqref{ray-bg} and \eqref{ray-pert} do, becoming respectively
\be
\f{a''}{a} + 4 \mH^2 = \f{4 \pi G \b\pi_\phi^2}{a^4} + \f{2 \b\pi_\phi^2}{3 a^7} \sum_j V_j \b E_j,
\ee
\be
\psi_n'' + 12 \mH \psi_n' = -\f{4 \pi G \b\pi_\phi^2}{a^4} \Big[ 8 \psi_n + 2(\de\pi_\phi)_n \Big]
- \f{2 \b\pi_\phi^2}{3 a^7} \sum_j V_j \b E_j \Big( 11 \psi_n + 2(\de\pi_\phi)_n
+ \f{\de E_{n,j}}{\b E_j} \Big),
\ee
Note that, in both equations, the last term involving the sum over $j$ disappears if $E_{n,j} = 0$.

It is also interesting that the presence of a bounce due to quantum gravity corrections is not obvious from these two equations. Indeed, at least when the matter content is a massless scalar field, the important quantum gravity corrections in the equations of motion for $a''/a$ and $\mH^2$, which are in the end responsible for the occurrence of the cosmic bounce, exactly cancel out in the sum $a''/a + 4 \mH^2$.  Note that the presence of the bounce can nonetheless be seen in \eqref{fried-bg} due to the negative term in the square root: the bounce happens when the term in the square root vanishes.

\subsection{Single-Spin Condensate States}
\label{ss.jo}

The equations of motion, for both background and perturbative degrees of freedom, simplify even further for single-spin condensates, i.e., for condensate wave functions $\sigma_{n,j}(\phi)$ that are non-vanishing for only one value $j_o$.  Note that this case can be reached dynamically in an asymptotic fashion for some GFT actions \cite{Gielen:2016uft, Pithis:2016wzf, Pithis:2016cxg}.

In addition, in the homogeneous case the resulting Friedmann equations for single-spin condensates are very similar to those derived in loop quantum cosmology (being of exactly the same form when $E_{n,j}=0$) \cite{Oriti:2016qtz, Oriti:2016ueo}, so this is a particularly interesting case to consider.

For a single-spin condensate, the correct classical limit is obtained for $m_{j_o}^2 = 3 \pi G$, and $(\pi_\phi)_n = \hbar \, Q_{n,j_o}$, so
\be
\b\pi_\phi = \hbar \, \b Q_{j_o}, \qquad (\de\pi_\phi)_n = \hbar \,\, \de Q_{n,j_o},
\ee
and $V_n = V_{j_o} \rho_{n,j_o}^2$.

The background equations \eqref{fried-bg} and \eqref{ray-bg} become, respectively,
\be \label{single-fried}
\mH^2 = \f{8 \pi G}{3} a^2 \ve \left( 1 - \f{\ve}{\ve_c} \right) + \f{8 V_{j_o} \b E_{j_o}}{9 a} \, \ve,
\ee
where $\ve = \b\pi_\phi^2/2 a^6$ is the energy density of the massless scalar field and $\ve_c = 3 \pi G \hbar^2 / 2 V_{j_o}^2 \sim 1/j^3 G^2 \hbar$ is the critical energy density, and
\be \label{single-ray}
\f{a''}{a} + 4 \mH^2 = 4 \pi G a^2 \ve + \f{2 V_{j_o} \b E_{j_o}}{3 a} \, \ve.
\ee
When $\b E_{j_o} = 0$, the bounce occurs when $\ve = \ve_c$, as can easily be seen in \eqref{single-fried}; in this case the emergent Friedmann equations agree exactly with the effective Friedmann equations of loop quantum cosmology (here given in terms of conformal time).

The equations of motion for the perturbations coming from \eqref{fried-pert} and \eqref{ray-pert} are:
\be \label{single-pert1}
\mH \psi_n' + \mH^2 \psi_n = - \f{4 \pi G \b\pi_\phi^2}{3 a^4} \Omega \Big[ 3 \psi_n + (\de\pi_\phi)_n \Big] - \f{2 V_{j_o} \bar E_{j_o} \b\pi_\phi^2}{9 a^7} \left[ 9 \psi_n + 2 (\de\pi_\phi)_n + \f{\delta E_{n,j_o}}{\bar E_{j_o}} \right],
\ee
\be \label{single-pert2}
\psi_n'' + 12 \mH \psi_n' = -\f{4 \pi G \b\pi_\phi^2}{a^4} \Big[ 8 \psi_n + 2(\de\pi_\phi)_n \Big]
- \f{2 V_{j_o} \b E_{j_o} \b\pi_\phi^2}{3 a^7} \left[ 11 \psi_n + 2(\de\pi_\phi)_n
+ \f{\de E_{n,j_o}}{\b E_{j_o}} \right],
\ee
where $\Omega = 1 - 2 \ve / \ve_c$.  These two equations are not independent: the second can also be obtained by differentiating \eqref{single-pert1} and using \eqref{pert-phi} as well as the equations of motion for the background.  (Similarly, for the background the relation \eqref{single-ray} can be obtained by differentiating \eqref{single-fried}, using $\b\pi_\phi' = 0$.)

Also, in the case $E_{n,j_o} = 0$, \eqref{single-pert1} and \eqref{single-pert2} are equivalent to the equations found for LQC using the separate universe approach \cite{Wilson-Ewing:2015sfx} (this is not surprising since these equations are derived from the Friedmann equations, which are identical for single-spin GFT condensates and in LQC).  And, as already mentioned for the more general case, for any value of $E_{n,j_o}$, in the classical limit these equations reduce to the standard equations of general relativity for long-wavelength scalar perturbations in the longitudinal gauge.

\

Since both $(\delta \pi_\phi)_n$ and $\delta E_{n,j_o}$ are constants of motion, either of the (equivalent) differential equations \eqref{single-pert1} and \eqref{single-pert2} can be solved to find how $\psi_n(t)$ evolves in time, given some initial conditions, once the background dynamics are known.  These initial conditions include cosmological observables like $(\de\pi_\phi)_n$ and $\psi_n(t_o)$, but also $\de E_{n,j_o}$ that come directly from the condensate wave function.  Therefore, unless $E_{n,j_o} = 0$, even for the simplest type of GFT condensate state it is impossible to obtain equations of motion for cosmological observables that depend only on cosmological observables: there remains an explicit dependence on $E_{n,j_o}$ (although this dependence becomes negligible in the classical limit).  Thus, the details of the GFT state that are unrelated to the main geometric and matter observables of interest nonetheless appear in the dynamics, showing that the dynamics of general relativity are modified by quantum features that do not appear to have an immediate geometric interpretation. While this is counterintuitive from the standard perspective on relativistic cosmology, it is natural to expect such a dependence from the quantum gravity point of view adopted here and more precisely from the perspective on cosmology as quantum gravity hydrodynamics, underlying our approach to cosmological perturbations as well. In hydrodynamics, in fact, it is to be expected that the emergent equations for specific macroscopic observables (which characterize the fluid only partially) do not close exactly, and retain a dependence on the underlying hydrodynamic variables (like density and velocity of the fluid) that are not macroscopic observables.

Finally, note that for a scalar field with some potential (which would be of particular interest for inflation), the equation of motion will be more complicated still, since in that case $(\delta \pi_\phi)_n$ will not be a constant of the motion.

\section{Discussion and Outlook}
\label{s.disc}

In this paper we have opened a new avenue towards the study of cosmological perturbations in a fundamental quantum gravity formalism by setting up a separate universe framework for long-wavelength scalar perturbations within group field theory condensate cosmology.

This was achieved by generalizing the simple type of GFT condensate quantum states put forward for homogeneous space-times to a kind of multi-condensate state, understood as a special case of the group field theory mean field states where quantum geometric data are localized in a relational sense by means of scalar fields used as clock and rods.

By using relational rods to locate the `homogeneous patches' of the separate universe framework that, put together, constitute the inhomogeneous space-time, it is possible to define the relevant GFT observables that, for the condensate states considered here, correspond to emergent cosmological quantities, whether background or perturbative.  To compare these cosmological observables in different patches, it was necessary to provide an appropriate notion of simultaneity in this background independent context.  The dynamics for the homogeneous background space-time and for the cosmological perturbations then emerge from the fundamental quantum equations of motion derived from GFT.  Finally, while we have not explicitly solved the resulting equations, we have shown how they reduce to the correct classical limit, as well as exhibited two special cases of interest for some GFT actions in which they simplify considerably.  These results demonstrate the viability of extracting effective continuum physics---and potentially testable predictions---from group field theory condensate cosmology.


The next research steps will move in two main directions: towards phenomenology, and further developing the framework.

First, an important goal is to make contact between the predictions of quantum gravity effects, here considered in the context of GFT, and observations of the early universe, particularly the cosmic microwave background.  This can be done by solving the emergent equations of motion derived here in contexts of interest, and calculating the resulting power spectrum which can then be compared to observations.  For example, in the matter bounce scenario the perturbations of interest, during the bounce, have a large wavelength compared with the Hubble radius so the equations derived here would be sufficient (if the dominant matter field during the bounce is a massless, or at least kinetic-dominated, scalar field).  On the other hand, for the inflationary scenario it is necessary to consider short-wavelength perturbations, and for ekpyrosis the scalar field cannot be approximated as massless; in both cases it will be necessary to extend the results derived here in order to study GFT-predicted effects in such a model.  (Note that in principle any matter field can be included in GFT, and the quantum dynamics will of course depend on the matter fields present.  Therefore, observable consequences will not necessarily be universal, but may depend, perhaps strongly, on the model being considered.)  That said, it may be possible to obtain an approximately scale invariant power spectrum directly from the GFT quantum gravity formalism without introducing any additional inflaton-like matter field.  Indeed, at least in some cases the quantum gravity dynamics alone generate a long-lasted accelerated expansion of the early universe \cite{deCesare:2016rsf}, without any need for an inflaton field.  Perhaps the inflaton can be replaced by a quantum gravity effect?

Second, it is necessary to extend the results obtained here in at least three directions: (i) understand how to treat short-wavelength perturbations, (ii) include tensor perturbations, and (iii) allow for a scalar field with a non-vanishing potential.  The last step may be the simplest, since the general form for the GFT action for gravity minimally coupled to a scalar field with any potential $V(\phi)$ is known \cite{Li:2017uao}.  However, even this case will be challenging as it will require working with a generalized GFT action, and in the emergent equations of motion for cosmological observables the scalar field momentum $\pi_\phi$ will no longer be a constant of the motion; as a result more work will be needed to arrive at a single equation of motion for the perturbations.

Concerning short-wavelength modes, it appears likely that, following \cite{Gielen:2017eco}, it will be necessary to include more terms in the GFT action, and in particular derivatives with respect to the scalar fields used as relational rods, which will play the role of spatial derivatives in the effective equations for cosmological perturbations.  Also, if one continues to work with a GFT state of the form \eqref{separatewave} with the condensate wave function constant in each patch (with respect to the relational rods), it will be necessary to include interactions between neighbouring patches (based on the interaction term in the GFT action).  More generally, it would be interesting to understand how the GFT interaction term can affect cosmological perturbations and whether it could leave an imprint in the power spectrum.

Finally, it is important to understand how to include tensor perturbations in GFT condensate cosmology.  The GFT observables considered so far are based on the volume (in each patch) and can be mapped directly to terms in the metric of a flat FLRW space-time with small perturbations expressed in the longitudinal gauge.  This is an isotropic observable (in the sense that it doesn't pick out any preferred direction with respect to the relational rods), and it seems likely that quantities that appear in off-diagonal terms in the metric will be related to non-isotropic GFT observables.  In fact, non-isotropic observables will likely be important for a number of observables beyond tensor modes.  For example, it would be nice to understand how to handle scalar perturbations in any gauge (or to introduce gauge-invariant variables like the Mukhanov-Sasaki variable), whereas here we worked in the longitudinal gauge.  To do this, it is necessary to allow the metric to have non-vanishing off-diagonal terms, for which non-isotropic GFT observables, as argued above, are likely relevant.  Also, even if the metric is diagonal, it can be important to pick out preferred directions as in the case of, e.g., the Bianchi I space-time.  This is particularly important for the early universe where, based on the Friedmann equation, it is expected that anisotropies will become important.  This will require going beyond the first step of including microscopic anisotropic configurations completed in \cite{deCesare:2017ynn}, and once again non-isotropic GFT observables will likely play an important role.  In short, a better understanding of non-isotropic GFT observables is important not only to extend GFT condensate cosmology to include tensor modes and to handle scalar modes in a gauge-invariant fashion, but also so anisotropies can be included as well.

\newpage

\raggedright

\end{document}